\documentclass[aps,prl,twocolumn]{revtex4}

\usepackage{amssymb}
\usepackage{amsmath}
\usepackage{epsfig}

\begin{document}

\title{Spatial solitons rays in periodic optical lattices}
\author { R. Khomeriki${}^{1,2}$,  J. Leon${}^1$}
\affiliation {
(${\ }^1$) Laboratoire de Physique Th\'eorique et Astroparticules \\
  CNRS-IN2P3-UMR5207, Universit\'e Montpellier 2, 34095 Montpellier (France)\\
(${\ }^2$)  Physics Department, Tbilisi State University, 0128
  Tbilisi (Georgia)}
\begin{abstract}
The light ray of a spatial soliton in an optical film whose refractive index is
smoothly modulated (wavelength much larger than the typical soliton width) in
both spatial directions is shown to possess chaotic regimes for which the
propagation is erratic. This is interpreted as a parametric driven pendulum,
obtained by a new perturbative approach of the Maxwell equation. These findings
are then demonstrated to compare well to the eikonal law of light ray
propagation (nonlinearity compensates diffraction).
\end{abstract}
\maketitle

The spatial optical solitons, that result from a balance of nonlinearity and
diffraction, manifest as stable self-focused light rays
\cite{kivsharBook,segevTrap,segevref,demetriTrap}. In a Kerr medium for
instance, the paradigm model obtained from Maxwell's equations is the nonlinear
Schr\"odinger (NLS) equation where the spatial direction of propagation plays a
role of \textit{``time''}. Thus many standard classical and quantum temporal
effects (as interaction, switching, tunnelling, etc.) can be usefully
experimented on permanent regimes in the spatial domain
\cite{silberbergColl,krolikColl,snyderColl,segevColl}.

Recent works consider various spatial localization scenarios in one and two
dimensional optical lattices
\cite{kevrekidisGrating,mandelikGap,ledererGrating}, where the soliton
localization width is comparable to (or even larger than) the intrinsic period
of the grating. Much less studies have been devoted to the spatial soliton
dynamics in \textit{smoothly inhomogeneous} media, although a
\textit{transversally} modulated refractive index has been studied in
\cite{russian,sivan}.
\begin{figure}[ht]
\centerline  {\epsfig{file=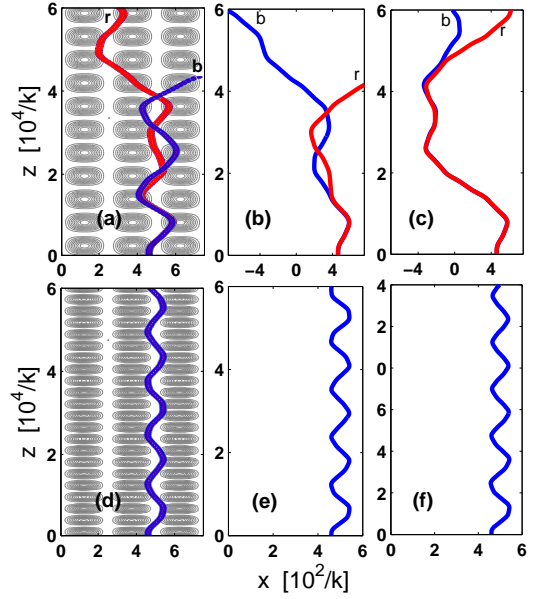,width=0.8\linewidth}}
\caption{Graph (a): contour plots of the electric field envelope
${\cal E}(x,z)$ solution of NLS (\ref{NLS}) in the potential
(\ref{land}) with initial conditions (\ref{ini}) for two slightly
different values of the injection point $x_0=460$ for the red (r)
ray and $x_0=461$ for the blue (b) one. The refractive index
landscape (\ref{land}) is plotted for reference. Graph (b) is the
driven pendulum approximation (\ref{acc33}) and graph (c)is the
Eikonal model (\ref{snell}), both simulated with the same initial
conditions ($A=2$ in all of those three cases). The lower graphs
(d), (e) and (f) describe corresponding simulations with same
initial conditions and parameters except for the longitudinal
modulation $\Omega$ replaced with $2\Omega$ (then here $A=0.5$).}
\label{fig:geo}\end{figure}
Our purpose here is to present the first discovery of erratic light rays that
occur in a smoothly periodic optical lattice obtained as in
\cite{kivshar111} by modulation of the refractive index in the two
spatial directions. Moreover the modulation along the direction of propagation
will be assumed smoother than the transverse one leading then to a quasi
two-dimensional problem.

Then we propose a new approach to the spatial soliton description where the
effect of lattice grating is considered as a two-wave nonresonant interaction.
We show that the chaotic motion is simply related to the parametric driven
pendulum which possesses periodic and chaotic regimes, instances of which are
presented in fig.\ref{fig:geo} and obtained by changing the longitudinal
modulation length of the optical lattice. Last but not least, this behavior is
demonstrated to follow the rules of light ray optics (eikonal) which actually
applies thanks to compensation of diffraction by nonlinearity.

The Maxwell's equations in a non-magnetic film belonging to the
$(x,z)$-plane is written hereafter for the out of plane linearly
polarized electric field component $E(x,z,t)$ by assuming
instantaneous response of the Kerr medium of susceptibility
$\chi$:
\begin{equation}\label{max}
\left(\frac {\partial^2 }{\partial x^2}+\frac
{\partial^2}{\partial z^2} \right) E - \frac1{c^2}\frac
{\partial^2}{\partial t^2} \left\{n^2E+\chi E^3\right\}=0.
\end{equation}
The linear refractive index  variations in the plane result from
the optical lattice, thus $n=n(x,z)$. The  $z$-axis is chosen
along the incident propagation direction and the standard approach
to spatial solitons then consists in assuming slowly varying
envelope, namely
\begin{equation}\label{def}
E=\sqrt{2/3}\,{\cal E}(\varepsilon x, \varepsilon^2 z) e^{i[\omega
t-kz]}+c.c. +{\cal O}(\varepsilon^2).
\end{equation}
with wave number $k=n_{_0}\omega/c$, where $n_{_0}$ is a reference refractive
index. The small expansion parameter $\varepsilon$ follows from the smallness of
the nonlinear Kerr coefficient, thus one may simply define $\chi=\varepsilon^2$.
Moreover we
are interested here in the physical situation when the spatial variations of the
refractive index are also weak, namely  $n_{_0}-n(x,z)={\cal O}(\varepsilon^2)$.
In such a case the Maxwell equation (\ref{max}) in the leading order reduces to
the NLS equation ($\varepsilon$ is scaled off)
\begin{equation}\label{NLS}
-i\frac {\partial{\cal E}}{\partial z}+\frac1{2}
\frac{\partial^2{\cal E}}{\partial x^2} +\chi|{\cal E}|^2{\cal E}
-\frac{n_{_0}-n(x,z)}{n_{_0}}{\cal E}=0,
\end{equation}
which is written in dimensionless space variables by the use of new  coordinates
$z\to kz$ and $x\to kx$. By assuming  different scalings along the
propagation direction $z$ and transverse direction $x$, the grating index
actually defines a quasi two-dimensional lattice, leading then to the above
spatial 1+1 dimensional equation in the external potential $1-n/n_{_0}$.

So far nothing has been assumed for the relative dimensions of the optical
lattice wavelength v.s. spatial soliton extension. As a matter of fact preceding
works as \cite{kevrekidisGrating,mandelikGap,ledererGrating} consider the case
when the spatial soliton \textit{sees} at least a few periods of the lattice
while here we are interested in the case when the lattice modulation
\textit{varies slowly} with respect to the typical soliton size.

To set an example let us assume a lattice periodic in both directions
\begin{equation}\label{land}
n(x,z)=n_{_0}\big[1-\chi V_{_0}\sin^2(Kx)
\sin^2(\Omega z)\big].
\end{equation}
with $0<\chi V_{_0}\ll 1$ such that $n_{_0}-n(x,z)$ is always positive.
Then we choose initial conditions for \eqref{NLS} representing the
injection of a laser beam in the $z$-direction,
\begin{equation}\label{ini}
{\cal E}(x,0)=a\,\textrm{sech}\left[a\sqrt{\chi}\ (x-x_0)\right]
\end{equation}
obtained from the exact soliton solution of (\ref{NLS}) for $n=n_{_0}$.
The fig.\ref{fig:geo}(a) now displays two numerical simulations of  (\ref{NLS})
with
\begin{equation}
 a=0.9,\quad \chi=0.01,\quad V_{_0}=0.2,
\quad  K=\sqrt{\chi}\pi/25,
\quad \Omega= K\sqrt{\chi V_{_0}},
\end{equation}
and two very close initial injection points $x_0=460$ (red ray) and $x_0=461$
(blue ray). The two spatial soliton rays are plotted against the potential
$1-n/n_{_0}$ for reference. Therefore two very close initial positions result in
very different trajectories indicating a chaotic nature of the soliton rays
(note the trajectories destabilize although $x_0$ is in a region where the
external potential vanishes). If now we increase $\Omega$, keeping all other
parameters values, we obtain a regime of a periodic stable trajectory presented
in fig.\ref{fig:geo}(d).

We propose hereafter a comprehensive interpretation of such a behavior in terms
of the equation of the parametric driven pendulum. Our approach is to consider
the Maxwell equation (\ref{max}) where the light beam interacts with the
index variations as a \textit{fundamental long-wave short-wave nonresonant
interaction process}. To that end we reconsider the reductive expansion method
by following \cite{oikawa} and seek a solution as
\begin{align}
& E=\varepsilon\sqrt{\frac{2}{3\chi}}\,\Psi(\xi_1,\zeta)
e^{i\theta(\xi_2,\zeta)}
e^{i[\omega t-kz]}+c.c.+\cdots,\nonumber\\
&(n(x,z)-n_{_0})/n_{_0}=\varepsilon^2V(\xi_2,\zeta)
\nonumber \\
&\xi_1=\varepsilon\left[x-\phi(\xi_2,\zeta)\right], \quad
\xi_2=\varepsilon^2x, \quad \zeta=\varepsilon^2 z. \label{multi}
\end{align}
The property that the index variations are smooth with respect to the spatial
soliton size is contained in the specific dependences on the slow variable
$\xi_1$ and very slow variable $\xi_2$.
The resulting equation at order $\varepsilon^3$ appears with terms
that depend solely on a single variable,  $\Psi(\xi_1,\zeta)$ on
the one side, $\theta(\xi_2,\zeta)$ and $V(\xi_2,\zeta)$ on the
other side. These terms thus decouple to eventually give
\begin{equation}\label{NLSor}\
-i\frac{\partial \Psi}{\partial\zeta}+ \frac{1}{2}\frac{\partial^2
\Psi}{\partial\xi_1^2}+\big|\Psi\big|^2\Psi=0,\quad
\frac{\partial\theta}{\partial\zeta}=-V.
\end{equation}
It appears therefore that the effect of the optical lattice, through
$V(\xi_2,\zeta)$, is not an external potential as in (\ref{NLS}) but is
contained in the coordinate drift $\phi(\xi_2,\zeta)$ defining $\xi_1$.
Considering now (\ref{max}) at order $\varepsilon^4$ we get
\begin{equation}\label{fin}
\frac{\partial\phi}{\partial\zeta}=-\frac{\partial\theta}{\partial\xi_2}.
\end{equation}
We eliminate first the phase $\theta$ between (\ref{NLSor}) and (\ref{fin}), and
second we turn back to initial spatial variables $x$ and $z$ to get
\begin{equation}\label{phi}
\frac{\partial^2\phi}{\partial z^2}= \frac{1}{n_{_0}}
\frac{\partial n(x,z)}{\partial x} .
\end{equation}
Then the nature of equation (\ref{NLSor}) allows us to start with an explicit
spatial soliton solution  (in physical dimensions)
\begin{equation}
\Psi_s=ae^{-\frac{i}{2}a^2\chi
z}\,\textrm{sech}\left[a\sqrt{\chi}\,(x-\phi)\right]
\end{equation}
whose position $x_s(z)$ is thus given by
$x_s-\phi(x_s,z)={\rm const}$. To compute the
\textit{``acceleration''} $d^2x_s/dz^2$ we must remember that
$\phi$ is slowly varying in $x$ and $z$ and thus keep only the
dominant orders, thus
\begin{equation}\label{acce-sol}
\frac{d^{\,2}x_s}{dz^2}\simeq
\frac{\partial^2\phi(x_s,z)}{\partial z^2}=
\frac{1}{n_{_0}}\frac{\partial n(x_s,z)}{\partial x_s} ,
\end{equation}
where the last relation follows from (\ref{phi}). This is the Newton law in the
\textit{``time''} $z$ for a particle of unit mass and position $x_s$ in the
\textit{time}-dependent potential $-n(x_s,z)/n_{_0}$.

Considering now the example (\ref{land}) used for the simulation in
fig.\ref{fig:geo}, the trajectory (\ref{acce-sol}) becomes the following
parametrically driven pendulum model \cite{pend1,pend2}
\begin{equation}\label{acc33}
\frac{d^{\,2}q}{dt^2}= -A\sin^2(t)\,\sin(q), \quad
A=2\chi V_{_0}\frac{K^2}{\Omega^2}
\end{equation}
where we have introduced a time $t=\Omega z$ and angle $q(t)=2Kx_s(z)$. These
reduced units show that the main control parameter is the above defined driving
amplitude $A$. The parametric instability of the pendulum with initial angle
$q(0)=0$ and velocity $\dot{q}(0)=0$ occur in the range $4/3<A<4$, while the
instability fastest growing rate is reached for
$A=16[\left(2/\sqrt{3}\right)\,-1]$. However for different initial conditions,
chaotic behavior may appear at smaller (or larger) values of $A$
\cite{pend1,pend2}. The graphs (b) and (e) in fig.\ref{fig:geo} show the result
of simulations of (\ref{acc33}) with $A=2$ and $A=0.5$, respecively for initial
velocity $\dot q(0)=0$ and initial angles corresponding to the rescaled
positions of the injected beams. The two different regimes presented result
simply from a modification of the modulation period of the lattice along the
propagation direction $z$. We observe indeed qualitative agreement, although
trajectories should not be quantitatively compared since the regime is chaotic.

In such solitonic regimes, nonlinearity compensates diffraction and the
resulting light ray is a good candidate to geometric optics. As a matter of fact
we demonstrate that the obtained soliton light ray equation (\ref{acce-sol}) can
also be viewed as the eikonal of a light ray in a medium with
variable index $n(x,z)$. The eikonal equation written for the position $\vec
r=(x,z)$ in the plane reads \cite{born-wolf}
\begin{equation}\label{eikonal}
\frac{d}{ds}\left(n\frac{d \vec r}{ds}\right)=\vec\nabla n,
\end{equation}
where $s$ is the arc length, namely $ds^2=dx^2+dz^2$. We write this equation for
the ray $x=x(z)$ as we did for the soliton motion, and obain after some algebra
that the above two components reduce to ($x'=dx/dz$)
\begin{equation}\label{snell}
x''=[1+(x')^2]
\frac{1}{n}\,\left(\frac{\partial n}{\partial x}
- x'\,\frac{\partial n}{\partial z}\right),
\end{equation}
where the right hand side  is eveluated on the trajectory $x=x(z)$. To recover
the soliton light ray, we must now use the assumption $x'\ll 1$, which is
precisely the assumption made to obtain the NLS equation (\ref{NLS}) from
Maxwell equation (\ref{max}). Taking also into account that $n(x,z)-n_{_0}\ll
n_{_0}$ one readily obtains from (\ref{snell})
\begin{equation}\label{snell-red}
x''\simeq\frac{1}{n_{_0}}\,\frac{\partial n}{\partial x},
\end{equation}
which is just the trajectory equation (\ref{acce-sol}). Numerical simulations of
(\ref{snell}) have been performed for the same parameter values as precedingly
(initial postion $x_0$, vanishing initial veleocities and explicit expression of
$n(x(z),z)$). The result is plotted in fig.\ref{fig:geo}(c) and
fig.\ref{fig:geo}(f), which confirms consistency with prediction of geometric
optics.

It is remarkable that such different models as NLS \eqref{NLS} in the external
potential smoothly modulated in both spatial directions and the driven
parametric pendulum \eqref{acc33} concur to describe trajectory of a spatial
soliton in a smoothly modulated (1+1) dimensional lattice. It is also striking
that the soliton ray actually follows the rule of geometric optics in this
context, which opens interesting perspectives on both theoretical and
experimental aspects.

One practical advantage of the parametric pendulum description is the prediction
of the switching from regular to chaotic (erratic) trajectories of the spatial
soliton. This result might be useful for conceiving all optical ultrasensitive
noise amplifiers. Last, our result is expected to apply to the case of
Bose-Einstein condensate with attractive nonlinearity in smooth space and time
dependent optical lattices, for which a similar dynamics will occur but there in
the time domain.

We are grateful to D. Felbacq and B. Guizal for enlighting dicussions. Work done
under contract CNRS GDR-3073. R.K. aknowledges stay as invited professor at the
\textit{Laboratoire de Physique Th\'eorique et Astroparticules} and financial
support of the Georgian National Science Foundation (Grant No GNSF/STO7/4-197).

\end{document}